\documentclass[11pt]{article}
\usepackage{amsmath}
\usepackage{amssymb}
\usepackage{amsfonts}
\usepackage{latexsym}

\def\beq{\begin{eqnarray}}
\def\eeq{\end{eqnarray}}
\def\ln{\,\mbox{ln}\,}

\def\al{\alpha}
\def\be{\beta}

\def\de{\delta}
\def\vp{\varepsilon}
\def\ep{\epsilon}

\def\ka{\kappa}
\def\la{\lambda}
\def\na{\nabla}
\def\pa{\partial}

\def\rh{\rho}
\def\si{\sigma}
\def\om{\omega}
\def\ph{\varphi}
\def\ta{\tau}

\def\Ga{\Gamma}
\def\De{\Delta}
\def\La{\Lambda}

\begin{document}

\begin{center}
{\Large Anomaly-induced effective action and Chern-Simons
modification of general relativity}
\vskip 6mm

{\bf Sebasti\~{a}o Mauro}$^{a}$ \ \
and
\ \
{\bf Ilya L. Shapiro}$^{b,a,c}$ \ \
\vskip 4mm

{\it (a)} \ \ Departamento de F\'{\i}sica, ICE,
Universidade Federal de Juiz de Fora,
\\
CEP: 36036-330, Juiz de Fora,
MG, Brazil
\vskip 4mm

{\it (b)} \ \
D\'epartement de Physique Th\'eorique and Center for
Astroparticle Physics, Universit\'e de Gen\`eve,
24 quai Ansermet, CH–1211 Gen\'eve 4, Switzerland
\vskip 4mm

{\it (c)} \ \
Tomsk State Pedagogical University and Tomsk State
University, Tomsk, 634041, Russia
\end{center}
\vskip 8mm

\begin{quotation}
\noindent
{\bf Abstract.}
\ \
Recently it was shown that the quantum vacuum effects of massless
chiral fermion field in curved space-time leads to the parity-violating
Pontryagin density term, which appears in the trace anomaly with
imaginary coefficient. In the present work the anomaly-induced
effective action with the parity-violating term is derived. The result
is similar to the Chern-Simons modified general relativity, which was
extensively studied in the last decade, but with the kinetic terms for
the scalar different from those considered previously in the literature.
The parity-breaking term makes no effect on the zero-order cosmology,
but it is expected to be relevant in the black hole solutions and in
the cosmological perturbations, especially gravitational waves.
\vskip 4mm

\noindent
Pacs: 04.62.+v, \ 	
11.10.Lm,       \ 	
11.15.Kc 	        
\vskip 2mm

\end{quotation}



\section{Introduction}

The derivation and properties of conformal (trace) anomaly are pretty
well-known (see, e.g., \cite{duff94}  and also \cite{PoImpo,PoS-Conform}
for the technical introduction related to the present work). At the
one-loop level the anomaly is given by an algebraic sum of the
contributions of massless conformal invariant fields of spins
$0,1/2,1$ in a curved space-time of an arbitrary background metric.
Recently, it was confirmed that the quantum effects of chiral (L)
fermion produce an imaginary contribution which violates parity
\cite{bonora}. As a result, the anomalous trace has the form
\beq
\langle T^\mu_\mu \rangle
&=&
-\,\be_1C^2
- \be_2E_4
- a^\prime{\Box}R
- \tilde{\be} F_{\mu\nu}^2
- \be_4 P_4\,.
\label{T}
\eeq
Here we have included the external electromagnetic field
$F_{\mu\nu}=\pa_\mu A_\nu-\pa_\nu A_\mu$ for generality, also
\beq
C^2
&=&
C_{\mu\nu\al\be}\,C^{\mu\nu\al\be}\,
\,=\,
R_{\mu\nu\al\be}^2 - 2 R_{\al\be}^2 + \frac13\,R^2
\label{Weyl}
\eeq
is the square of the Weyl tensor in four-dimensional space-time
and
\beq
E_4 &=& \frac14\,\vp^{\mu\nu\al\be}\,\vp^{\rho\si\la\tau}
\,R_{\mu\nu\rho\si}\,
R_{\al\be\la\tau}
\,=\,
R_{\mu\nu\al\be}^2 - 4 R_{\al\be}^2 + R^2
\label{GB}
\eeq
is the integrand of the Gauss-Bonnet topological term.

The $\be$-functions are given by algebraic sums of the
contributions of $N_s$ scalars, $N_f$ Dirac fermions and
$N_v$ massless vector fields. The explicit form is well
known,
\beq
(4\pi)^2\,\be_1
 &=& \frac{1}{120}\,N_s + \frac{1}{20}\,N_f +
\frac{1}{10}\,N_v\,,
\nonumber
\\
(4\pi)^2\,\be_2 &=& -\,\frac{1}{360}\,N_s
- \frac{11}{360}\,N_f - \frac{31}{180}\,N_v\,,
\nonumber
\\
(4\pi)^2\,\be_3 &=& \frac{1}{180}\,N_s
+ \frac{1}{30}\,N_f - \frac{1}{10}\,N_v\,.
\label{abc}
\eeq
One can assume that $a^\prime$ in (\ref{T}) is equal to $\be_3$,
but there is ambiguity, as will be discussed below.
${\tilde \be}$ is the usual $\be$-function of QED or scalar QED etc,
depending on the model.

Furthermore, there is a parity-violating Pontryagin density term
$\be_4 P_4$, where
\beq
P_4
&=&
\frac12\,\vp^{\mu\nu\al\be}\,R_{\mu\nu\rho\si}\,
R_{\al\be}\,^{\rho\si}\,.
\label{Pont}
\eeq
By dimensional reasons the term with $P_4$ is possible, but for a
long time it was believed that this term, in fact, does not show
up. However, in a recent paper \cite{bonora} this term was
actually found with a purely imaginary coefficient
$\,\be_4 = i/(48 \cdot 16\pi^2)$, as a contribution of chiral
(left) fermions. The chirality is important
here, because the contribution of the right-hand fermions is going
to cancel the one of the left-hand fermions, so taking them in a
pair would kill the effect. Let us also note that much earlier, in
\cite{DuffNieuw}, the possibility of such a term coming from
integrating out antisymmetric tensor field has been considered,
also some general considerations were presented even earlier in
\cite{DDI-80} and more recently in \cite{Nakayama}.

Some questions arise due to the result of \cite{bonora} and
its physical interpretation. First, does the parity-violating term
in the anomaly mean that the dynamics of gravity is affected in a
significant way? Second, in case of a positive answer to the last
question, does it mean that the chiral fermions are disfavoured
theoretically, since they produce imaginary component in the
gravitational field equations? The last possibility was discussed
in \cite{bonora} as a theoretical argument in favor of massive
neutrino. The third question is whether the parity-odd
terms in the anomaly have some relation to the Chern-Simons
modification of $4d$-gravity suggested in \cite{jackiw-pi,LWK}.
The theories of this sort were extensively investigated in the
last decade, as one can see from the review \cite{AleYu} and
other works on the subject. This question looks really natural,
because the Chern-Simons-gravity is based on the action which
includes the $P_4$-term with an extra scalar factor inside the
integral. Let us note that the relation between parity-odd terms
and anomalies in $D=4$ was discussed, i.e., in \cite{AlvaWit} in
relation to gravitational anomalies, so the novelty of the
term (\ref{Pont}) concerns only the trace anomaly.

The purpose of the present work is to address the questions
formulated above.
In order to do so, we derive the effective action of gravity by
integrating conformal anomaly, and show that the result is a new
version of the Chern-Simons $4d$-gravity with a special form of the
kinetic term for the scalar and some extra higher-derivative
terms which are typical for this action. From the technical side
most of the consideration is pretty well-known, but we present
full details in order to make it readable for those who are not
familiar with the subject. The paper is organized as
follows. In Sect. 2 we review the well-known scheme of deriving
anomaly-induced effective action, with an extra parity-odd term
corresponding to Pontryagin density. The anomaly-induced action
provides a specific form of the kinetic term for the auxiliary
scalar in Chern-Simons modified gravity. For this reason, in
the last subsection we present a short review of the previous
version of kinetic terms, which are known in the literature.
Sect. 3 includes a general, mainly qualitative, discussion of
the physical interpretation of the new parity-violating term.
Finally, in Sect. 4 we draw our conclusions and suggest
possible perspectives of a further work on the subject.

\section{Integration of anomaly with parity-violating term}

The integration of conformal anomaly (\ref{T})  in $d=4$ means solving
the equation similar to the one for the Polyakov action in $d=2$,
\beq
\frac{2}{\sqrt{-g}}\,g_{\mu\nu}
\,\frac{\de\, {\bar \Ga}_{ind}}{\de g_{\mu\nu}}
\,=\,
-\,\langle T_\mu^\mu \rangle
\,=\,
\frac{1}{(4\pi)^2}\,
\big(\,\om C^2 + bE_4 + c{\Box} R + \tilde{b} F_{\mu\nu}^2
+ \ep P_4\,\big)\,.
\label{mainequation}
\eeq
Here we introduced useful notations
$\,\big(\om,\,b,\,c,\,{\tilde b},\,\ep\big)
= (4 \pi)^2\,\big(\be_1,\,\be_2,\,a^\prime,\,\tilde{\be},\,\be_4\big)$.
The coefficient $\ep$ derived in \cite{bonora} is imaginary, but we
will not pay attention to this until the solution is found.
The first reason for this is that this is technically irrelevant,
and also it is, in principle, possible to have a real coefficient
of the same sort at the non-perturbative level.

\subsection{Conformal properties of Pontryagin term and anomaly}

The solution of Eq. (\ref{mainequation}) is technically
is not very complicated \cite{rei} in the usual theory without
Pontryagin term, and it remains equally simple when this term
is present. In order to understand this, let us make an observation
that this term is conformal invariant in $d=4$, simply because one
can recast (\ref{Pont}) in the form when the Weyl tensor replaces
the Riemann tensor,
\beq
P_4
&=&
\frac12\,\vp^{\mu\nu\al\be}\,C_{\mu\nu\rho\si}\,
C_{\al\be}\,^{\rho\si}\,.
\label{Pont-W}
\eeq
The proof of this statement is well-known (see \cite{Gru-Yu}
for further developments), but for the convenience of the reader
we present a proof in the Appendix.
One can easily see that the {\it r.h.s.} of the Eq.
(\ref{mainequation}) consists of the three different terms,
which can be classified according to \cite{DeserSchwim}. One
can distinguish {\it (i)} conformally invariant part
$\om C^2 + \tilde{\be} F_{\mu\nu}^2 + \be_4 P_4$; \ {\it (ii)} the
topological term $bE_4$ and {\it (iii)} surface term $c\Box R$.

In fact, the last division is not unambiguous. For example, in $d=4$
both $P_4$ and $bE_4$ can be presented as total derivatives, and the
term $P_4$ is not only topological, but also conformal, according
to Eq. (\ref{Pont-W}). Hence, the Gauss-Bonnet invariant can be
attributed to two groups of terms and the Pontryagin density even
to all three groups {\it (i)}, {\it (ii)} and {\it (iii)}. In any
case, as the reader will see shortly, the conformal invariance
of $P_4$ makes the inclusion of this term into anomaly-induced
action a very simple exercise. We shall present some details only
to achieve a self-consistent exposition of the consideration.

The simplest part is the $\Box R$-term, which can be directly
integrated by using the relation
\beq
- \,\frac{2}{\sqrt{-g}}\,g_{\mu\nu}
\,\frac{\delta }{\delta g_{\mu\nu}}\,\int d^4x\sqrt{-g}\,R^2
\,=\, 12\,{\square} R\,.
\label{identity}
\eeq
It is easy to see that in this case the solution is a local
functional, that gives rise to the well-known ambiguity in the
coefficient $a^\prime$ of the
$\Box R$-term, which was discussed in details in \cite{anom2003}.

Now, let us concentrate on the non-local part of anomaly-induced
action\footnote{The non-localities due to anomaly was first
discussed in \cite{DDI-76}.}.
The solution of (\ref{mainequation}) can be presented in
the simplest, non-covariant form, in the covariant non-local form
and in the local covariant form with two auxiliary fields. Let
us start from the simplest case. By introducing the conformal
parametrization of the metric
\beq
g_{\mu\nu} &=& {\bar g}_{\mu\nu}\,e^{2\si(x)}
\label{confpa}
\eeq
one can use an identity
\beq
- \frac{2}{\sqrt{-g}}\,g_{\mu\nu}
\frac{\de\,A[g_{\mu\nu}]}{\de\, g_{\mu\nu}}
= - \frac{1}{\sqrt{-{\bar g}}}\,
\left.\frac{\de\,A[{\bar g}_{\mu\nu}\,e^{2\si}]}{\de \si}
\,\right|_{{\bar g_{\mu\nu}}\rightarrow g_{\mu\nu},
\si\rightarrow 0}\,.
\label{deriv}
\eeq
Here and below the quantities with bars are constructed using
the metric ${\bar g}_{\mu\nu}$, in particular
\beq
{\bar F}_{\mu\nu}^2
&=&
F_{\mu\nu}\,F_{\al\be}\,
{\bar g}^{\mu\al}{\bar g}^{\be\nu}\,,
\qquad
F_{\mu\nu} = \pa_\mu A_\nu - \pa_\nu A_\mu
\label{Fmn}
\eeq

Furthermore, we will need the conformal transformation rules
\beq
\sqrt{-g}\,W_k &=& \sqrt{-{\bar g}}{\bar W}_k^2
\,,\qquad
\mbox{where}
\qquad
\big(W_k = C^2,\,P_4,\,F^2\big)\,,
\label{Wk}
\eeq
and
\beq
\sqrt{-g}(E - \frac23{\Box}R)
&=&
\sqrt{-{\bar g}}({\bar E}
- \frac23{\bar {\Box}}{\bar R} + 4{\bar {\De}}_4\si)\,,
\label{GBtrans}
\\
\sqrt{-{\bar g}}\,{\bar \De}_4 &=& \sqrt{- {g}}\,{\De}_4\,,
\nonumber
\eeq
where
\beq
\De_4
&=&
\Box^2 + 2R^{\mu\nu}\na_\mu\na_\nu
- \frac23\,R\Box +\frac13\,R_{;\mu}\,\na^\mu
\label{Pan}
\eeq
is covariant, self-adjoint, fourth-derivative, conformal
operator \cite{Paneitz}.

After we use the transformation rules (\ref{GBtrans}) and
(\ref{Wk}), Eq. (\ref{mainequation}) becomes very simple
and the solution for the effective action can be found in
the form
\beq
{\bar \Ga}_{ind}
&=&
\frac{1}{(4\pi)^2}\,
\int d^4 x\sqrt{-{\bar g}}\,\Big\{
\om \si {\bar C}^2 + \tilde{\be} \si {\bar F}_{\mu\nu}^2
+ \ep {\bar P}_4
+ b\si \big({\bar E}-\frac23\, {\bar {\bar \Box}} {\bar R}\big)
\,+\,
2b\si{\bar \De}_4\si\Big\}
\nonumber
\\
&-&
\frac{1}{12}\,\Big(c+\frac{2b}{3}\Big)
\frac{1}{(4\pi)^2}\,
\int d^4 x\sqrt{- g}\,R^2
\,+\,
S_c[{\bar g}_{\mu\nu},\,A_\mu]\,,
\label{quantum}
\eeq
where $S_c[{\bar g}_{\mu\nu},\,A_\mu]=S_c[g_{\mu\nu},\,A_\mu]$
is an unknown conformal invariant functional of the metric
and $\,A_\mu$. This functional is an integration constant for
the Eq. (\ref{mainequation}) and hence it can not be uniquely
defined in the present framework. Let us note that in some
cases this functional is irrelevant. An example is cosmological
solution without background
electromagnetic field. In this case the metric is conformally
trivial and $S_c[g_{\mu\nu}]$ becomes an irrelevant constant.

Even in cases of non-cosmological metrics the functional
$S_c[g_{\mu\nu}]$ does not prove to be very significant,
because the rest of the effective action (\ref{quantum}) contains
all information about the UV behaviour of the theory. In the
massless case, with a usual duality between UV and IR regimes,
this means that  $S_c[g_{\mu\nu}]$ may have only sub-leading
contributions. These arguments are confirmed by successful
applications to black holes \cite{balsan,Reis-Nord} and
gravitational waves \cite{wave,GW-HD}.

The solution (\ref{quantum}) is non-covariant, because it is
not expressed in terms of the physical metric $g_{\mu\nu}$.
In order to obtain the non-local covariant solution of Eq.
(\ref{mainequation}), one has to introduce the Green function
for the Paneitz operator,
\beq
\big(\sqrt{-g}\De_4\big)_x\,G(x,y) = \de(x,y)
\quad
\mbox{and notation}
\quad
\int_x = \int d^4 x \sqrt{-g (x)}
\,.
\label{GP}
\eeq
Using (\ref{deriv}) it is easy to check that for any conformal
functional $A(g_{\mu\nu}) = A({\bar g}_{\mu\nu})$,
\beq
&&
2 g_{\mu\nu} (y)\,\frac{\de}{\de g_{\mu\nu} (y)}
\,\int_x\,A\cdot\big(E - \frac23{\Box}R\big)
\label{gabi}
\\
&=&
\frac{\de}{\de \si (y)}\,\int_x\,A\cdot
\left.\big(E
- \frac23{\Box}R\big)\right|_{\si \to 0\,,\,\, {\bar g}_{\mu\nu} \to g_{\mu\nu}}
\,=\,
4\sqrt{-{\bar g}}{\bar {\De}}_4 \,A
\,=\, 4\sqrt{- g}{\De}_4 \,A\,.
\nonumber
\eeq

By means of the last relation it is easy to solve both remaining parts
(remember that the local part we already have from Eq. (\ref{identity}))
of induced effective action, and we arrive at
\beq
{\bar \Ga}_{ind}
&=& \Ga_{\om} + \Ga_b + \Ga_c\,,
\label{nonloc}
\eeq
where
\beq
\Gamma_\om \, = \,\frac14\,\int_x\,\int_y
\, \big(\om C^2 + \tilde{b} F_{\mu\nu}^2 + \ep P_4 \big)_x
\,G(x,y)\,\big(E - \frac23{\Box}R \big)_y\,,
\label{om-term}
\eeq
\beq
\Gamma_b = \frac{b}{8}\,\int_x\,
\int_y\,
\big(E - \frac23{\Box}R\big)_x\,G(x,y)\,\big(E - \frac23{\Box}R\big)_y
\label{b-term}
\eeq
and
\beq
\Ga_{c} = - \frac{c+\frac23\,b}{12(4\pi)^2}
\,\int_x\,R^2(x) \,.
\label{c-term}
\eeq
One has to note that the Pontryagin density shows up only in the
first nonlocal term (\ref{om-term}), but in what follows we shall
see that the second term (\ref{b-term}) is still relevant for
constructing the kinetic term of the Chern-Simons modification
of gravity. At the same time, the local term (\ref{c-term}) will
remain separated from others.

\subsection{Anomaly-induced action and kinetic term
for Chern-Simons gravity}

As a next step,
the nonlocal expressions for the anomaly-induced effective
action can be presented in a local form by introducing two auxiliary
scalar fields $\ph$ and $\psi$ \cite{a}. An equivalent two-scalar
representation was suggested in \cite{MaMo}, while the simpler
one-scalar form was known from much earlier \cite{rei}.
Since the details of the procedure were described also in
\cite{PoImpo,PoS-Conform} and do not change essentially due to
the term $P_4$, let us present only the final result
\beq
\Gamma_{ind}
 &=& S_c[g_{\mu\nu}]
- \frac{3c+2b}{36(4\pi)^2}\,\int_x\,R^2
+  \int_x\,\Big\{\frac12 \,\ph\De_4\ph - \frac12 \,\psi\De_ 4\psi
\nonumber
\\
&+& \ph\,\left[\,\frac{\sqrt{-b}}{8\pi}\,
\big(E -\frac23\,{\Box}R\big)\,
-\, \frac{1}{8\pi\sqrt{-b}}\,
\left(\om C^2 + \tilde{b} F_{\mu\nu}^2 + \ep P_4 \right)\,\right]
\nonumber
\\
&+&  \frac{1}{8\pi \sqrt{-b}}\,\psi\,\left(\om C^2
+ \tilde{b} F_{\mu\nu}^2 + \ep P_4\right) \,\Big\}\,.
\label{finaction}
\eeq
At the classical level the local covariant form (\ref{finaction}) is
equivalent to the non-local covariant form (\ref{nonloc}).
The definition of the boundary conditions for the Green
functions \ $G(x,y)$ are equivalent to the same boundary conditions
for the auxiliary scalars $\ph$ and $\psi$. For the discussion of
the importance to have two fields let us address the reader to
\cite{a,MaMo,PoImpo}.

The action (\ref{finaction}) represent a final product of our
integration of conformal anomaly. However, in order to make the
consideration leading to the version of Chern-Simons gravity
\cite{jackiw-pi} more explicit, let us make a change of variables
similar to one of \cite{MaMo}. Let us introduce two new scalars,
\beq
\chi = \frac{\psi - \ph}{\sqrt{2}}\,,
\qquad
\xi = \frac{\psi + \ph}{\sqrt{2}}\,,
\label{chixi}
\eeq
such that
\beq
\ph = \frac{\xi - \chi}{\sqrt{2}}\,,
\qquad
\psi = \frac{\chi + \xi}{\sqrt{2}}\,.
\label{phpsi}
\eeq
Then the total gravitational action, including the
classical part and the anomaly-induced action (\ref{finaction})
can be cast into the form
\beq
\Ga_{grav} &=&
S_{EH} + S_{HD} + \Gamma_{ind}
\nonumber
\\
&=&
S_{EH}[g_{\mu\nu}] + S_{HD}[g_{\mu\nu}] + S_c[g_{\mu\nu}]
\,+\,  \int_x\,\Big\{\xi\De_4\chi
\,+\,
k_1\big(E -\frac23\,{\Box}R\big)\big(\xi - \chi\big)
\nonumber
\\
&+&  k_2\,
\chi\,\left(\om C^2
+ \tilde{b} F_{\mu\nu}^2 + \ep P_4\right)
\,+\,k_3R^2
\Big\}\,.
\label{faction}
\eeq
where
\beq
k_1=\frac{1}{8\pi}\,\sqrt{-\frac{b}{2}}\,,
\qquad
k_2=\frac{1}{8\pi \sqrt{-2b}}\,,
\qquad
k_3=\,-\,\frac{2b+3c}{36\,(4\pi)^2}\,,
\label{coeffs}
\eeq
and the classical vacuum part includes the Einstein-Hilbert action with
cosmological constant
\beq
S_{EH}
\,=\,-\,\frac{1}{16\pi G}\int d^4 x\sqrt{-g}\,\left(\,R+2\La\,\right)\,.
\label{EH}
\eeq
and the higher derivative terms,
\beq
S_{HD} &=& \int d^4x \sqrt{-g}
\left\{a_1C^2+a_2E+a_3{\Box}R+a_4R^2 \right\}\,.
\label{HD}
\eeq
Obviously, the $R^2$-terms in (\ref{HD}) and (\ref{faction})
combine, that produce a well-known ambiguity in the local part
of the total action with anomaly-induced contribution \cite{anom2003}.

Compared to the previously known solutions
\cite{a,MaMo,GiaMot,QED-Form}, the expression (\ref{faction}) has an
extra term proportional to $\,\chi\,P_4$, where $\chi$ is a
new auxiliary scalar field related to the conformal anomaly. This
is exactly the structure which was extensively discussed in the
context of Chern-Simons extension of general relativity starting
from \cite{LWK} and \cite{jackiw-pi}. The remarkable difference
is that the field $\chi$ in (\ref{faction}) has higher-derivative
kinetic term and also contains a mixing with the second scalar
field $\xi$.

\subsection{Brief review of other forms of the kinetic term}

Since the main output of the previous consideration is the new
form of the kinetic term for the Chern-Simons modified gravity,
Eq. (\ref{faction}), it is worthwhile to give a list of the
previously known kinetic terms.

The Chern-Simons gravity is usually understood as an effective
theory which should be obtained from a more fundamental theory
\cite{jackiw-pi,LWK}. Consequently, the form of the kinetic
term depends on the choice of the fundamental theory. In our
case it is the quantum theory of matter fields (with parity
violation) on classical curved background, which led us to
(\ref{faction}). This action is different from the previously
known versions, which can be presented as
\beq
S &=& \int d^4x\sqrt{-g}\, \Big\{
-\ka R
+ \frac{\al}{4}\,\psi \tilde{R}^{\mu\nu\al\be} R_{\nu\mu\al\be}
- \frac{\be}{2}\,\big[\,g^{\mu\nu}\na_\mu\psi\na_\nu\psi
 + 2V(\psi)\,\big] \Big\} + S_{mat} \,,
\label{dCS}
\eeq
where $\ka=1/16\pi G$, while $\al$ and $\be$ are some new constants.
The Chern-Simons coupling field is $\psi$ with a potential term
$V(\psi)$, and $S_{mat}$ is the action of matter. Also,
we used the standard notation for the dual Riemann tensor
\beq
\tilde{R}^{\mu\nu\al\be}
&=&
\frac{1}{2}\, \ep^{\al\be\rh\si}\,
R^{\mu\nu}_{\,\,\,\,\,\,\,\, \rh\si} \,.
\eeq
The modified Enstein equation for the action (\ref{dCS}) is
\beq
R_{\mu\nu} - \frac{1}{2}\,g_{\mu\nu}R + \frac{\al}{\ka}\,
C_{\mu\nu} = \frac{1}{2\ka}\,T_{\mu\nu} \ ,
\label{EMmetrica}
\eeq
where $T_{\mu\nu}$ is the total momentum-energy tensor.
The C-tensor is defined as
\beq
C^{\mu\nu}
&=&
\na_{\al}\psi \,\ep^{\al\be\rh(\mu} \,\na_\rh R^{\nu)}_{\,\,\,\,\be}
\,+\, \tilde{R}^{\be(\mu\nu)\al}\,\na_\be\na_\al\psi \,.
\eeq
The variation of (\ref{dCS}) with respect to the scalar field yields
the equation
\beq
\Box\psi
&=&
\frac{dV(\psi)}{d\psi} - \frac{\al}{4\be}\,\tilde{R}^{\mu\nu\al\be}
\,R_{\nu\mu\al\be} \,,
\label{EMescalar}
\eeq
which is the Klein-Gordon equation with an extra Pontryagin
density source.

There are two main approached, based on different
choices of the constants $\al$ and $\be$. One of them is called
non-dynamical Chern-Simons gravity, when we have $\be=0$. Then
the scalar field does not evolve dynamically, and is a field
prescribed externally.
This model was introduced by Jackiw and Pi in \cite{jackiw-pi},
where it was defined that $\psi=t/\mu$, the choice called
canonical, with  $\mu$ being some dimensional parameter. The
boundary conditions in this theory were discussed in \cite{Gru-Yu}.
The development of this approach and further references can be
found in the review \cite{AleYu}.

All solutions for the non-dynamic case must satisfy the
Pontryagin constraint
\beq
\tilde{R}^{\mu\nu\al\be}
R_{\nu\mu\al\be} = 0 \ .
\eeq
This constraint limits the space of solutions of the theory. For
instance, Kerr metric can not be solution since it does not satisfy
this constraint. The rotating black hole solutions have been found
within approximation schemes, with certain inconsistencies discussed
in the literature \cite{yunes-pretorius}. On the other hand, it was
discussed that ghosts can not be avoided in the non-dynamic theory
\cite{MotoSuy}, forcing to pay more attention to the dynamic scalar
case.

The dynamic case corresponds to an arbitrary $\al$ and $\be$
in the action (\ref{dCS}). Such a model was introduced by Smith
et. al. in \cite{smith}, motivated by the low energy limit
of string theory. The potential term was supposed to follow
from the fundamental string theory, however, as usual, there
is some freedom in this part.
For a zero potential, $V(\psi)=0$, there is a
uniqueness theorem which ensures that in case spherically
symmetric, static and
asymptotically flat spacetime, the solution is given by the
Schwarzschild metric \cite{shiromizu}.  In \cite{rogatko}, the
uniqueness was established for the case the Reissner-Nordstrom
metrics.

Until now, there are no exact solutions for a rotating black
hole and some approximate schemes are used instead. For example,
in \cite{yunes-pretorius,kMT,YSYT2} the slowly rotating black
hole was studied in the small-coupling limit, while \cite{Ali-Chen}
carried out the study for an arbitrarily large coupling.

Recently, the post-GR corrections for the dynamic model has
been considered through the study of rapidly rotating black holes
in the decoupling limit \cite{stein,konno-takahashi}. Such work is
important because of the possibility to constraint the theory in
the strong field regime.

The gravitational perturbations are fundamental for better
understanding of the gravitational waves and the stability
of solutions. In \cite{MotoSuy,Yunes-Sopuerta,CG} the gravitational
perturbations were explored for the black hole background and
in \cite{DFK} for the cosmological case.
The issue of ghosts was studied for both non-dynamic and dynamic
models. Although in both models the ghosts are present, one can
avoid them for a constant scalar background in the dynamic case
\cite{MotoSuy}, while in the non-dynamical theory the scalar
behavior is fixed.
Let us mention that the works listed above were done for zero
potential of the scalar field, while in the recent work
\cite{MPCG} the gravitational perturbations with the mass term
$V(\psi)=m^2\psi^2$ were considered.

Indeed, the choice of the kinetic term is non-trivial, in
particular it was recognized that the Lagrangian of the
kinetic term does not necessarily be of the Klein-Gordon
type. Other kinetic terms were considered, e.g. the one
discussed in \cite{AleYu} has some relation to string theory
\beq
S_{\psi} = -\frac{1}{2} \int d^4x \sqrt{-g}\Big\{ \be_1\,
g^{\mu\nu}\na_\mu\psi\na_\nu\psi
\,+\,
\be_2\,(g^{\mu\nu}\na_\mu\psi\na_\nu\psi)^2 \Big\} \,,
\eeq
where $\be_1$ and $\be_2$ are some constants.

An approach which is the closest one to our result (\ref{faction})
was developed in \cite{YS,YSYT} and eventually used to describe
the linear stability in \cite{AYY}.
The corresponding theory is known as Quadratic Modified Gravity,
the action can be cast into the form
\beq
S &=& \int d^4x\sqrt{-g} \Big\{\ka R +  f_1(\psi)R^2
+ f_2(\psi)R_{\mu\nu}^2 + f_3(\psi)R_{\mu\nu\al\be}^2
\nonumber
\\
&+&
f_4(\psi)\tilde{R}^{\mu\nu\al\be}R_{\nu\mu\al\be}
- \frac{\be}{2}\,\Big[\,g^{\mu\nu}\na_\mu\psi\na_\nu\psi
+ 2V(\psi)\,\Big] \Big\} + S_{mat} \,,
\label{fi}
\eeq
where $f_i(\psi)$ are some functions of the scalar field.
One can easily see that in the case of (\ref{faction}) all
these functions are linear, the coefficients are defined
by the number of quantum particles and the kinetic term
is more complicated and involves higher derivative and the
second scalar. Also, the potential term is absent for the
anomaly-induced version of the Chern-Simons gravity
(\ref{faction}).

The situation with ghosts and tachyons in the theory (\ref{fi})
was discussed in \cite{MotoSuy}. It is important to note that
the fields $\chi$ and $\xi$ are auxiliary scalars, which just
exist to parametrize the non-localities in the original action
(\ref{nonloc}). This feature removes the need for discussing
the ghosts related to the fields $\chi$ and $\xi$. Another
way to understand this is to remember that the terms in the
action (\ref{nonloc}) are at least of the third order in
curvature (except $R^2$, which does not produce ghosts
\cite{Stelle-77}). Therefore, the quantum part of induced 
actions presented
above has no issue with ghosts, at least on the flat
background\footnote{This does not exclude the emergence of 
ghosts on other background, as it was discussed in \cite{Chiba-05}
and recently in \cite{QGprT} in relation to the final stage of 
the de~Sitter phase of the evolution of the $\La$CDM universe.}. 
Of course, the classical action behind the
anomaly contains a usual $C^2$-term, which is known to produce
ghosts. However, there are some indications that the ghost is
not becoming a real particle at the energies below Planck scale
\cite{GW-HDQG}.

\section{Interpretation and applications of the
parity-violating terms}

The presence of imaginary parity-violating terms in the
conformal anomaly (\ref{T}) can be interpreted such that
the existence of massless left-handed neutrino should be
theoretically disfavoured \cite{bonora}. This possibility
looks very interesting, especially in view of experimental
confirmation of neutrino oscillations. However, one has to
remember that the conformal anomaly is not a directly
observable physical quantity. The remarkable exception is the
cosmological FRW-like solution, when the metric depends only on
the conformal factor according to Eq. (\ref{confpa}), with
$\si=\si(\eta)$ and $\eta$ conformal time. The trace anomaly
directly affects the dynamics of $\si=\si(\eta)$. But, as
far as Weyl tensor is zero for the FRW-like metric, the Eq.
(\ref{Pcon}) shows that the Pontryagin term is also zero for
this metric. Consequently, the background cosmological solution
is not affected by the presence of the new term with $P_4$.

In all other cases the solution (\ref{faction}) is
not exact. This means that the effect of the  $P_4$-dependent
term can be, in principle, compensated by the conformal functional
$S_c(g_{\mu\nu})$. This means that all the conclusions concerning
the possible physical effects related to $P_4$ assume that the
functional $S_c(g_{\mu\nu})$ is irrelevant. On the other hand,
the general arguments presented above show that this assumption
is quite reasonable. Then, we can expect that the $P_4$-term can
be relevant for the gravitational waves on the cosmological (or
other) background, and also for the physically relevant solutions
such as Schwarzschild, Reissner-Nordstrom or Kerr.

The two aspects of the $P_4$-term in the anomaly (\ref{T}) and
action (\ref{faction}) can be relevant. The first one is that
this term is parity-violating. This means that it is expected
to produce a parity-odd solutions, including for the metric
perturbations. As a result, one can expect the parity-odd
component to emerge in the CMB spectrum. The second aspect is
related to the imaginary coefficient. Let us present some
considerations of these two aspects, but start from the
general discussion of the solution in the presence of the
Pontryagin term.

\subsection{Possible gravitational solutions}

The solutions in the Chern-Simons modified theory of gravity have
been extensively discussed in the literature, e.g., in the papers
\cite{YS,AYY}. The main difference between the models
which were previously considered and (\ref{faction}) is the
form of the kinetic term for the scalar field and the presence
of higher derivative terms. Let us consider in some details the
simplest case of the spherically-symmetric solution, which is
quite illustrative. Our purpose is not to find a new solution,
but only show that the parity-violating and other higher-derivative
terms in the action (\ref{faction}) may modify the
usual Schwarzschild solution. This does not happen with the
classical higher derivative terms of (\ref{HD}), because for the
Ricci-flat background the Weyl-square term in $d=4$ can be
easily reduced to the Gauss-Bonnet topological invariant
\cite{BH-dim}. So, we can completely concentrate on the
anomaly-induced part. The equations for the auxiliary fields
have the form
\beq
\De_4\xi &=& k_1 \big(E -\frac23\,{\Box}R\big)
- k_2 \left(\om C^2 + \ep P_4 \right) \,,
\nonumber
\\
\De_4\chi &=& - k_1 \big(E -\frac23\,{\Box}R\big)\,.
\label{cx}
\eeq
Furthermore, in the Ricci-flat case the Paneitz operator
becomes simply $\Box^2$, and also one has
$E=C^2=R_{\mu\nu\al\be}^2$. For the sake of simplicity, consider
the possible solutions of the form \cite{jackiw-pi}
\beq
\xi = d_1 t + f_1(r) \,,
\qquad
\chi = d_2 t + f_2(r) \,,
\label{CXsol1}
\eeq
where $d_{1,2}$ are constants and $f_{1,2}$ some functions of the
radial variable $r$. Assuming that the metric satisfies Schwarzschild
solution, one has $\,P_4=0$, $\,R_{\mu\nu\al\be}^2=48(GM)^2/r^6\,$ and
\beq
\Box^2f_{1,2}(r) = \frac{\al_{1,2}\,(GM)^2}{r^6}\,.
\label{emsb}
\eeq
where $\,\al_1 = 12(k_1-\om k_2)$ and $\,\al_2 = -12k_1$.
The general solution for the functions $f_{1,2}(r)$ corresponds
to the equations (here $f=f_{1,2}$ and $\al=\al_{1,2}$) was
obtained in \cite{balsan},
\beq
\frac{df}{dr}
&=& \frac{Br}{3} +\frac{2MB}{3}-\frac{A}{6}-\frac{\al}{72M}
\,+\, \left(\frac{4}{3}BM^2 + \frac{C}{2M}-AM -\frac{\al}{24}\right)
\,\frac{1}{r-2M}
\nonumber
\\
&-&
\frac{C}{2M}\frac{1}{r}
\,- \,\frac{\alpha M}{18}\,\frac{\ln r}{r(r-2M)}
\,-\, \left(\frac{A}{2M}-\frac{\alpha}{48M^2}\right)\,\frac{r^2\,\ln r}{3(r-2M)}
\nonumber
\\
&+&
\left(\frac{A}{2M}-\frac{\al}{48M^2}\right)
\,\frac{\left(r^3 -8M^3\right)\ln(r-2M)}{3r(r-2M)}\,.
\label{CXsol2}
\eeq
Here $(d,A,B,C)$ are constants that specify the homogeneous
solution of $\Box^2f=0$.
However, it is not necessary that the equation for the metric can
be satisfied for any choice of the coefficients $(d,A,B,C)_{1,2}$.
In the part which is important for our consideration,
however, one can see that there is no real influence of the $P_4$ term,
at this level. Moreover, if we assume, as an approximation, that at
low energies higher derivative parity-even terms are irrelevant and
only parity-odd $P_4$-dependent term is relevant, then the Schwarzschild
solution is valid in this truncated version of the theory. Finally, Eq.
(\ref{CXsol1}) which is typical \cite{balsan} in the higher derivative
model such as (\ref{faction}), shows that the difference between
dynamical and non-dynamical versions of the Chern-Simons modified
gravity can be resolved, at least for some particular solutions.

\subsection{Parity violation}

What could be the effect of parity-violating term in the
gravitational action? In order to answer this question, one has
to remember that the most likely manifestation of the Pontryagin
term is the parity violation in the gravitational waves spectrum
\cite{LWK}.
Due to the Planck suppression the effect is going to be very weak
and perhaps can not be observed directly. However, the parity
violation can eventually go to the CMB through the well-known
mechanisms (see, e.g., \cite{Dodelson,Durrer-CMB}) and may
eventually lead to the anisotropy in the metric perturbations.

\subsection{Imaginary coefficient}

The imaginary component of effective action is a typical phenomena
in quantum field theory \cite{Schwinger-51}. Usually it is related
to the logarithmic structure in the form factors at the UV (the
same as conformal anomaly) and indicates to the possible particle
production by external field \cite{ItsikZu,Maggiore}. In order to
have such a production, the energy of created particles should
be smaller than the intensity of an external field. In the case
of strictly massless neutrino this condition can be
easily satisfied. However, some other details must be taken
into account. The production of massless left-handed neutrino
will be related to the fourth-derivative term $\,k_2 \chi\ep P_4$
in the effective action (\ref{faction}). This term is
strongly suppressed by the Planck mass in the Einstein-Hilbert term,
even in the inflationary period, except in the initial stable phase
of the modified Starobinsky inflation model \cite{asta}. And in
this special case any kind of particle production is compensated
by the powerful inflation, such that the density of created
particles remains negligible.

After inflation the energy of the created neutrino particles
would be very small, at most of the order of the energy of the
gravitational waves, since the effect is zero for the FRW-like
background. During the long period of
existence of the Universe there can be certain production of such
particles, but there is another aspect of the problem. The
neutrinos are fermions and, with a very small energy, fermionic
particles should form a Fermi surface. Then the production of
neutrino should be suppressed by the Pauli principle.
From the quantum theory viewpoint this means the existence of
the effect of quantum interaction between neutrino, which would
forbid their creation. All in all, the imaginary component of
the effective action can not be probably seen as a basis of the
no-go theorem forbidding theoretically massless neutrino.

\section{Conclusions}

We presented a simple derivation of the anomaly-induced
effective action of gravity with the new parity-violating
term in conformal anomaly, which was recently discovered in
\cite{bonora}. The integration proceeds with a minimal
changes compared to the known procedure, since the new
term is both topological and conformal. The result of the
integration represents a new version of the well-known
Chern-Simons modification
of general relativity, which was extensively discussed in the
literature, starting from \cite{jackiw-pi} and \cite{LWK}.
The possible role of such a term was explored in details
\cite{AleYu}, and we can not add much to this discussion,
except to suggest a new form of kinetic term for the
auxiliary scalar field, derived in (\ref{faction}).

Concerning the physical significance of the Pontryagin term,
there is no doubt that the presence of parity-violating term
in gravity is potentially very interesting \cite{AleYu}. The
reason is that even a very small violation of the symmetry can
give an observable effect. At the same time, some simple
qualitative arguments show that the effect of imaginary
term in the effective action and the related production of
neutrino by the gravitational background should be too
weak to provide a theoretical ``prohibition'' of the
massless neutrino, as it was suggested in \cite{bonora}.

One more observation concerns the electromagnetic sector
of the anomaly-induced action. There is no parity violation
in this part of the action. However, if one could find some
field (in the baryonic or dark sectors of the spectrum),
which produce the parity-violating term in the conformal
anomaly, the mechanism which we described in this paper
would immediately generate axion with a very specific
form of the kinetic term, equal to the one presented in
(\ref{faction}).

Finally, let us say a few words about the perspectives of the
new form of the Chern-Simons modified gravity (\ref{faction}).
It is obvious that it would be interesting to check both
theoretical and phenomenological consequences of this theory,
starting from the solution for the rotating black hole and
cosmological applications. From the QFT side, it would be
interesting to see whether some (probably reduced) version
of this term can be derived in
other theories, in particular whether it can be met in the
theory on massive neutrino at very low energies, due to the
difference of the masses of the right and left components
from one side and the effect of gravitational decoupling
\cite{apco} from another side. Regardless of the serious
technical difficulties of this program, it does not look
completely impossible to be completed.

\section*{Acknowledgements}

I.Sh wish to acknowledge stimulating correspondence and
discussions with Dr. Guilherme Pimentel, clarifying conversation
with Dr. Stefano Giaccari and also interesting discussion about
related subjects with Prof. Fidel Schaposnik.
S.M. is grateful to CAPES for support of his Ph.D. research program.
I.Sh. is very grateful to the D\'epartement de Physique Th\'eorique
and Center for Astroparticle Physics of Universit\'e de Gen\`eve
for partial support and kind hospitality during his sabbatical
stay, and to CNPq and also to FAPEMIG and ICTP for partial
support of his work.

\section*{Appendix}

The Pontryagin density is given by
\beq
P_4 &=& \frac{1}{2}\ep^{\mu\nu\al\be}
R_{\al\be \, . \, .}^{\,\,\,\,\,\,\,\, \ta \la}
R_{\mu\nu\ta\la} \,.
\label{P4}
\eeq
Let us prove that the Riemann tensor here can be replaced by the
Weyl tensor. In 4-dimensional space we have the following relation
between Riemann and Weyl tensors,
\beq
R_{\mu\nu\al\be} = C_{\mu\nu\al\be} + \frac{1}{2}
( R_{\mu\al} g_{\nu\be} - R_{\mu\be} g_{\nu\al}
+ R_{\nu\be} g_{\mu\al} - R_{\nu\al} g_{\mu\be})
- \frac{1}{6} R ( g_{\mu\al} g_{\nu\be}
- g_{\mu\be} g_{\nu\al}) \,.
\label{R-W}
\eeq
Replacing (\ref{R-W}) into (\ref{P4}) we get
\beq
P_4 = \frac{1}{2}\ep^{\mu\nu\al\be}
C_{\al\be \, . \, .}^{\,\,\,\,\,\,\,\, \ta \la} C_{\mu\nu\ta\la}
+ 2 \ep^{\mu\nu\al\be}C_{\mu\nu\la\be} R^\la_\al
- \frac{1}{3} \ep^{\mu\nu\al\be} C_{\mu\nu\al\be} R \,.
\eeq
On the other hand, Bianchi identity for the Weyl tensor
\beq
C_{\mu\nu\al\be} + C_{\mu\be\nu\al}
+ C_{\mu\al\be\nu} = 0\,,
\eeq
provide the relations
\beq
\ep^{\mu\nu\al\be}C_{\mu\nu\la\be} = 0 \ \ , \ \
\ep^{\mu\nu\al\be} C_{\mu\nu\al\be} = 0 \,.
\eeq
Therefore, we arrive at the desired formula,
\beq
P_4 =  \frac{1}{2}\ep^{\mu\nu\al\be}
C_{\al\be \, . \, .}^{\,\,\,\,\,\,\,\, \ta \la}
C_{\mu\nu\ta\la} \,,
\label{Pcon}
\eeq
that shows $\sqrt{-g} P_4$ to be conformal invariant.


\end{document}